# Development of a low flow vapor phase electrochemical reactor without catholyte reliance for $CO_2$ electrolysis to high value carbon products

Ashley Cronk[a,b], David Larson[a], and Francesca M. Toma[a]

a Chemical Sciences Division and Joint Center for Artificial Photosynthesis, Lawrence Berkeley National Laboratory, Berkeley, CA

b Department of Civil & Environmental Engineering, University of California, Berkeley, Berkeley, CA

**Abstract:**
A continuous $CO_2$ vapor-fed electrochemical cell prototype that performs $CO_2R$ with high faradaic efficiency for desired carbon products for up to 72 hours of operation is presented. The cell design facilitates a flow through configuration, capitalizing on gas diffusion electrode (GDE) and membrane electrode architecture (MEA) without requiring a catholyte. We demonstrate stable performance and design adaptability by incorporating various $CO_2R$ catalysts and membranes into the cell, investigating lifetime performance and selectivity under established experimental conditions. With an Ag foil and PPO anion exchange membrane, the design achieves an average current density of -30mA/cm$^2$ and FE ~80% for CO obtained over a 12-hour duration. With a Cu-based catalyst, FE ~40% selectivity for ethylene was achieved. The unique geometry and flexibility of the cell provides an adaptive and dynamic GDE electrochemical cell framework, easing future research for novel electro-catalysts on GDE electrodes and MEA/GDE assemblies for enhanced (photo)electrochemical carbon dioxide reduction.

**Introduction:**
Future energy consumption projections coupled with the adverse effects of climate change have sparked efforts towards developing sustainable alternatives to current energy generation practices. Artificial photosynthesis has gained profound interest as an avenue to achieve carbon neutrality by (photo)electrochemically reducing carbon dioxide into valuable synthetic fuels and chemicals.[1] Bocarsly et al. coins the phrase "solar fuels based on reverse combustion," conveying the concept of using carbon dioxide as a feedstock for sustainable organic fuel generation.[2] In order to understand the feasibility of (photo)electrochemical reduction of $CO_2$, efforts are currently concentrated on pure electrochemical environments that facilitate the most efficient and selective system without being limited by the required cell potential deficiency. Products of electrochemically reduced carbon dioxide vary from a wide array of hydrocarbons (i.e. methane, ethane, and ethylene), to $H_2$, and CO. These synthetic fuels possess various uses, serving as an important intermediate for the Fischer-Tropsch reactions, more complex carbon-based fuels, and syngas.[3-5] Catalyst selection determines the distribution of carbon species produced via electrochemical reduction. Ag and Au exhibit the highest known selectivity towards CO, while Cu and corresponding oxide species generate diverse carbon products, thus acquiring the most attention as a novel candidate for artificial photosynthesis.[6]

The most common electrochemical cell system for carbon dioxide reduction has been traditionally in aqueous environments.[7] In aqueous systems, gas phase $CO_2$ attempts to saturate the catholyte while electrodes, typically planar in structure, are submerged in electrolyte. This environment introduces diffusion limitations, hindering efficiency and $CO_2$ feedstock concentration. The main barrier is the relatively low solubility of $CO_2$. Calculated Henry's constants from various gaseous species demonstrates $CO_2$ difficulty saturating solutions relative to its partial pressure (0.034 M). Because $CO_2$ exhibits low solubility in aqueous solutions, $CO_2$ concentrations are limited at the electrode-catalyst surface.[7] As a result, the low solubility of $CO_2$ hinders generated current densities due to mass transfer limitations from the mass-transfer boundary layer thickness at the electrode.[8] Theory and experimental modeling suggest that vapor-fed $CO_2$ electrochemical systems present a promising alternative for combating mass transport limitations and enhancing triple phase boundary reactions (electrochemically active areas in the cell) for efficient and selective $CO_2R$. To combat mass transfer limitations and increase $CO_2$ concentration at the electrode-catalyst surface, integrating gas diffusion electrodes (GDE) have been identified as a promising alternative to aqueous cell systems, enabling vapor phase $CO_2$ to be fed directly into the cathode. Traditional applications of GDEs spawned from fuel cell applications. Substantial characterization work has been done by A. El-kharouf et al. who provided a comprehensive survey of GDE materials and corresponding properties, noting that carbon-based GDE are the most common.[13]

Several studies establish that GDE structures generate higher current densities than planar electrodes due to the increase of catalytically active sites inherit from their porous structure.[8-11] Multi-walled carbon nanotube AgNP-based GDEs has been shown to achieve current densities of up to 350mA/cm$^2$.[9] GDEs fabricated from electrodeposited Sn on carbon fibers have been reported to increase catalytic activity, generating formate at a FE of 71% for a lifetime of 6 hours.[11] Work has also been invested investigating catalytic reaction rates in GDE systems dependent on the $CO_2$ flow configuration in relation to the GDE-catalyst surface. For example, Xiang and co-workers report a flow-through GDE system that generates sufficiently higher current densities than a flow-by system both in aqueous media. This system was utilized for high-rate electrochemical reduction of carbon monoxide and it was demonstrated that the flow-through GDE system enhanced reaction rates of the Cu-nanoparticle catalyst.[10]

An electrochemical system that facilitates direct vapor-fed $CO_2$ eliminates the dependency of acidic/basic electrolytes and allows the integration of sensitive catalysts, photo-absorbers, and other semiconductor thin-films into the cell system. In highly acidic electrochemical systems that rely on proton-exchange, the competing hydrogen evolution reaction (HER) is promoted due to the availability $H^+$ (hydrogen) and $H_2O^+$ (hydronium) ions in the cathode. If HER is favored, faradaic efficiencies (FE) decline for reduced carbon products. To suppress the competing HER, alkaline electrolytes and anion exchange membranes have been employed in several $CO_2R$ electrochemical systems.[10,15,16] However, it must be noted in flow through configurations, a truly alkaline catholyte isn't feasible due to carbonate equilibrium chemistry since dissolved $CO_2$ creates carbonic acid, resulting in neutral to slightly acidic conditions in water.

For electrochemical reduction of $CO_2$ to achieve economic feasibility and future implementation, focus should be extended from not only GDE-catalyst synthesis but advanced characterization through custom electrochemical cell design. Many studies that focus on GDE-catalyst synthesis use commercially available electrochemical flow cells for catalyst characterization.[11,14] Survey work of varying reactor designs for $CO_2$ electrochemical flow cells was conducted by Enrodi et al, providing an extensive study of cell configurations capitalizing on GDE architecture and their corresponding components. The most universal cell design described includes vapor $CO_2$ fed passively on the other side of the catholyte. This system combats most of the limitations of traditional $CO_2$ reduction but encounters challenges with industrial scale-up due to pressure sensitivity.[6] In addition, the use of a catholyte in this design is still present. Thus, the development of practical reactor designs that operate while vapor $CO_2$ is fed directly into the cathode, has the potential to overcome performance and solubility issues[20] and increase the feasibility of industrial scale-up.

Even with commercially available reactors, efforts to optimize the electrochemical system using a "filter-press" cell was shown, where experimental work was conducted of tuning liquid and gas flows of the system, however the system was not fully vapor-fed.[11] Irabien et al. presents a "filter-press" electrochemical membrane reactor without the use of a catholyte but achieved current densities that oscillated around 7.5mA/cm² with a 24% deviation for 100 minutes while achieving FE for $CH_4$ ranging from 5 to 12%.[14] Salvatore et al. presents a humidified $CO_2$ vapor-fed electrolyzer flow cell with a bipolar membrane to accommodate the oxygen evolution reaction and discrepancies between pH during electrochemical reduction. However, FE stability for CO declined after 1 hour of operation likely attributed to the de-wetting of the catalyst surface.[16] A similar gas-fed $CO_2$ electrolyzer was presented two years later, where experimental work compared FE and current densities with the addition of a $NaHCO_3$ support layer in the cathode. Initial experiments were conducted in a fully vapor-fed system, but FE stability relied on a $NaHCO_3$ solid supported aqueous layer in the cathode to reach FE ~ 80% and current densities of 30mA/cm².[15] From this work, the reliance of hydration in the cathode for vapor-fed $CO_2$ electrochemical systems are identified as crucial parameters in their system design for achieving acceptable FE for CO production at high current densities.

The inquiry of catholyte reliance in electrochemical $CO_2$ reduction is elusive due to the variance of reported cell performances. Even though cathode hydration is identified as a crucial factor in cell efficiency and $CO_2$ product selectivity, developing a system that operates with its absence can decrease cell resistance due to the decreased distance between the electrodes and membrane assembly. Zero-gap device configurations have been presented but their shortcomings are enabled by their enhanced ability to achieve extremely high current densities.[6]

In this study, we present a continuous vapor-fed $CO_2R$ electrochemical flow reactor design that dismisses catholyte reliance, overcoming hydration and stability barriers presented by previous similar works.[14,15,16] We also experimentally demonstrate stable operation for up to 72 hours and achieve carbon product selectivity (FE ~ 80%) for 12 hours. In addition, 40% FE for ethylene is achieved with a Cu catalyst. The main objective of this report is to demonstrate the stability, longevity, and selectivity of our cell design using Ag and Cu electrocatalyst GDEs. In this report, we study the advantages of a CO2 vapor fed system and the use of GDE-based electrocatalysts in both the anode and cathode of the cell.

**Experimental:**

*Cell design and fabrication:* The final cell design is a product of various influences stemming from electrochemical membrane electrode assembly (MEA) electrolyzers and single-stacked fuel cell devices. Cell designs have been presented with similar external geometry and internal flow field configurations for solar water splitting.[19] The overall design concept resembles a zero-gap device where internal cell components are pressed together with the exclusion of any internal aqueous supports. The only aqueous component is fed into the anode as batch anolyte. The main difference between our device geometry and devices presented prior comes from the accommodation of gas diffusion and membrane electrode assembly, lack of catholyte, and use of GDEs in both the anode and cathode of the cell.

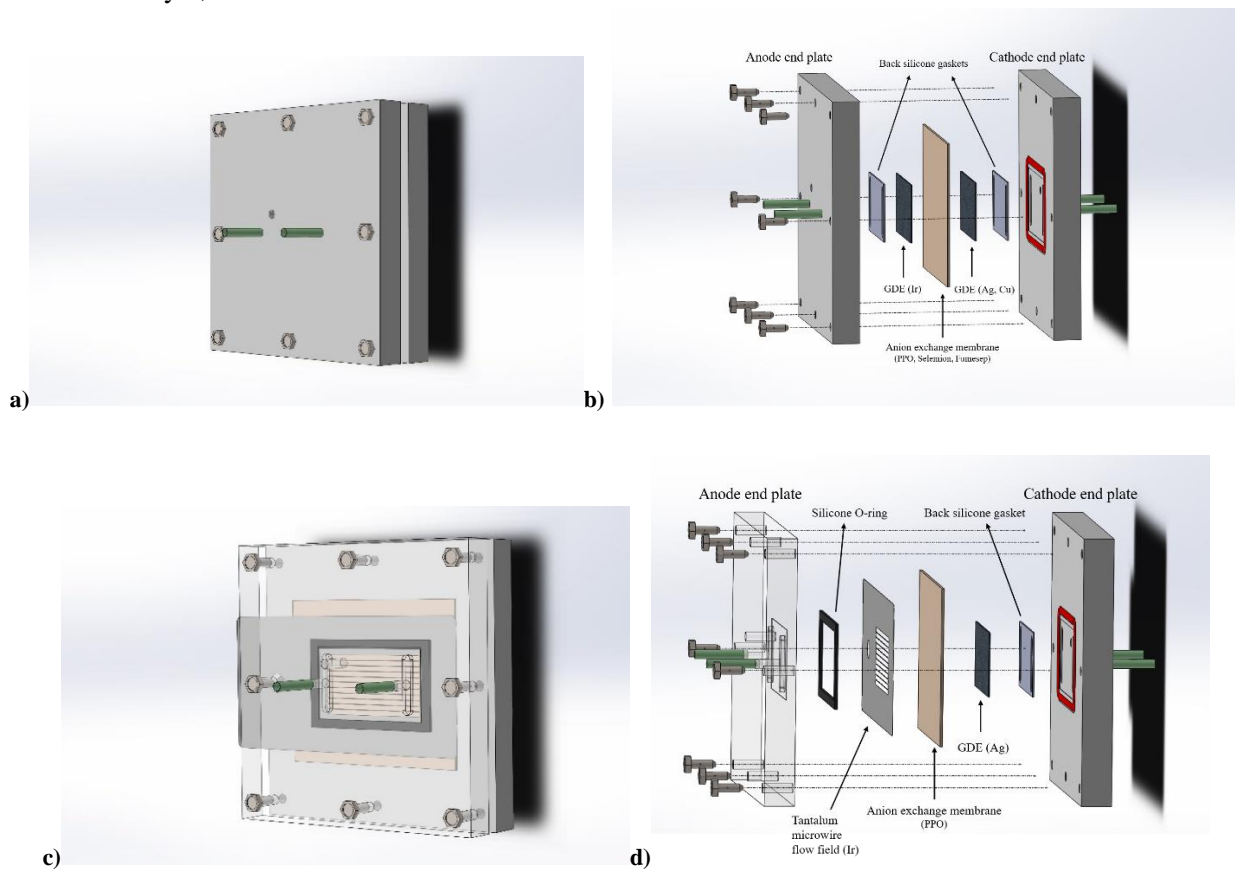

**Figure 1.** Test bed electrochemical cell schematic with components. a) Front view (anode) of electrochemical flow cell with GDE architecture in the anode and cathode. b) Corresponding exploded view schematic. c) Front view (anode) of microwire anode-photocathode flow cell. d) Corresponding exploded view schematic.

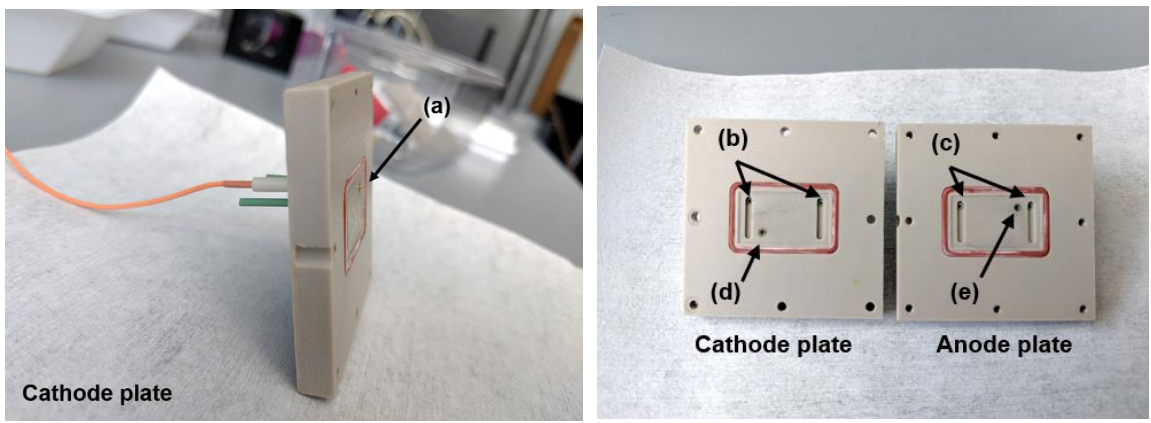

**Figure 2.** Electrochemical cell design with GDE architecture in both the anode and cathode. (a) spring loaded Au wire as the cathode current collector. (b) reactant and product flow ports. (c) flow ports for anolyte. (d) current collector port for spring loaded Au wire. (e) current collector port for spring loaded Pt wire.

As shown in Figure 1, the cell is constructed in a two-electrode configuration where the anode and cathode are parallel. For both configurations shown, the anode (front) and cathode (back) plates are structurally identical (2''x 2''x 0.25''). For the dark GDE configuration, both anode and cathode plates are machined out of Polyether ether Ketone (PEEK), a chemically resistant high-performance thermoplastic polymer. For the anode microwire-photocathode configuration, the anode plate is machined out of transparent polycarbonate. Each plate possesses flow ports (PEEK tubing) for reactant and product transport and electrical connectivity. Flow ports are connected perpendicular with two-part epoxy (EPO-TEK 302-3M) and cured for 24 hours. Internally, both the cathode and anode plate possess a centrally located 100-micron recession to accommodate the GDEs in addition to the anion exchange membrane and various gaskets for reactant/product separation (Figure 2). The recession depth was determined after several design iterations, identifying the optimal depth to ensure proper contact and electrical connectivity. Surrounding the recession, a silicone O-ring was installed to seal the compartment. The internal cell compartment and flow port tubing location was designed to facilitate a reactant flow-through system. This allows gaseous $CO_2$ to flow-through the GDE-catalyst component while forcing direct contact with the catalyst layer, membrane, and anodic (oxidized) products. A silicone back gasket was installed between the GDE and cell plate to prevent any gaseous $CO_2$ to bypass the GDE.

Electrical connectivity is accomplished using a spring-loaded Pt wire in the anode and an Au wire in the cathode (Figure 2). These spring-loaded wires are inserted through the silicone back gasket and placed such that it rests between the silicone back gasket and back electrode surface. The Au wire is being used instead of Pt in the cathode to suppress hydrogen production.

The overall internal cell volume was designed to be minimal (< 0.026 cm$^3$) with the intention of reducing dead space with no electrode. We hoped that low flow rates (~5sccm of CO2 or lower) would ensure sufficient opportunity for CO2 reduction to occur before product analysis to maximize utilization efficiency. Even though most work with GDE architecture in vapor-fed systems operate with $CO_2$ flow rates ranging from 17sccm to 100sccm,[15,18,22] low flow rates of $CO_2$ emulate more practical situations where $CO_2$ concentrations may be dilute and obtained from natural or point sources (i.e. carbon emissions and the atmosphere).[20] Low flow rates through the anode flow-field facilitates non-turbulent flow and decreased instability. High batch anolyte flow rates have been tested through our system and was found to introduce bubbles, physical instability, and catalyst delamination. Low flow rate conditions require less energy and are more practical for industrial scale-up.

*Membranes:* To achieve alkaline conditions, anion (OH$^-$) exchange membranes were used. For Ag-catalyst experiments, a PPO membrane (poly (2,6-dimethyl-1,4-phenylene oxide) suspended in Methanol) or Selemion AMV anion exchange membrane (AGC Engineering Co., LTD.) was used. Selemion membranes were measured to have a thickness of 110µm. PPO membranes were cast with a target thickness of 96-100µm. Due to the brittle nature of PPO, membranes are hydrated in millipore water after overnight curing for eased membrane handling and installation. Lastly, an industrial grade anion exchange membrane,

Fumasep FAA-50 with a thickness of 40μm, was used for Cu-catalyst experiments. Fumasep membranes are hydrated in batch anolyte solution (0.05M K2CO$_3$) overnight before use.

*Electrocatalyst fabrication:* The working electrode substrates are constructed from woven carbon paper (Toray 060, wet proofed, PTFE treated) cut with a surface area of 2.7cm$^2$. Prior to depositions, the carbon substrates are rinsed in 10% nitric acid solution to eliminate metal contaminants, specifically in the cathode. Two different working electrodes were fabricated for electrochemical experiments. The Ag electrocatalysts are fabricated by depositing a 20nm thick Ag foil, RF sputtered at 100W in a 3mTorr Ar environment. The Cu electrocatalysts are fabricated via electrodeposition from a highly basic (~13 pH) copper sulfate precursor by applying a current density of -0.4mA/cm$^2$ until uniform coverage (~5 hours). Substrates were plasma etched prior to depositions to increase catalyst penetration during electrodeposition. Optimal deposition times were determined iteratively such that uniform coverage was obtained to maximize catalyst surface area and penetration into the substrate. These morphologies were confirmed by SEM. For final cell design electrochemical experiments, the counter electrode is a 20nm thick Ir foil, RF sputtered at 100W in 3mTorr Ar environment on woven carbon paper substrate (Toray 060, wet proofed, non-PTFE treated). For GDE architecture comparison in the anode, a tantalum microwire flow plate with sputtered Ir was used (Figure 1b). These anode flow fields were laser cut to a 100μm thickness. The flow fields consist of channels 17mm in length, 1mm wide with the resulting width of the wire channels to be 100μm.[19] This configuration is similar to flow field electrodes used in prior works by Kistler et al.

*Cell assembly:* Prior to cell assembly and in between experimental testing, cell plates and components undergo extensive cleaning to ensure minimal catalyst cross-contamination during experimental operation. Cell plates and gaskets are soaked in Chelex solution (1:10 water) to replace any transition metals with sodium, then soaked in 10% nitric acid solution overnight. Tubing that transports products and reactants undergo similar procedure.

The cell is assembled using a mounting platform which allows components to be aligned and properly compressed during installation. Cell assembly begins with the cathode (back) plate. The mounting platform possesses an inlet that allows the cathode plate with flow ports to lay flush during assembly. The silicone back gasket is installed in the centrally located recession followed by the cathode GDE. The hydrated membrane is then placed on top, ensuring that the membrane extends past the silicone O-ring that seals the internal compartment. The anode (front) plate is assembled in a similar manner and placed on top of the cathode plate. Once aligned, the plates are compressed using eight screws that are tightened in an alternating fashion such that the internal components are compressed evenly.

*Device measurements and system set-up:* To evaluate cell design and outcomes of the GDE/MEA architecture, we performed extensive electrochemical measurements focusing on long term operation, performance stability, and selectivity. Electrochemical reactions and data were carried out and collected using a Biologic SP-200 potentiostat where the working electrode is connected to the cathode and the counter electrode is connected to the anode. For gaseous product analysis, the cathode output is connected to a gas chromatograph (SRI 8610C) that is calibrated to detect hydrogen, oxygen, carbon monoxide, methane, ethane, and ethylene. The gas chromatograph possesses two channels, one which is equipped with a flame ionization detector (FID) for $CO_2$ product detection. The cycle length for each injection is approximately 9 minutes with variable buffer time between measurements. Buffer times between gaseous product analysis were chosen depending on the nature of the measurement (*pre-operando* characterization or *in-operando* characterization). The system under operation conditions consists of a peristaltic pump (Cole Parmer Masterflex C/L) flowing 5sccm of 0.05M $KHCO_3$ anolyte, purged with nitrogen, through the anode chamber. Batch anolyte is used for each experimental run. On the cathode side, 5sccm of humidified $CO_2$ flows continuously through the cathode input using a mass flow controller (Alicat Scientific). Relative humidity (RH) is monitored in the cathode (Sensirion SHTW2), obtaining an average humidity for CO of 80% RH during operating conditions. All experiments occur in ambient conditions.

*In-operando* characterization occurs after 24 hours of membrane hydration within the cell. A linear sweep voltammetry (LSV) measurement is run with voltage limits set to -1.5V to -4V with a scan rate of 100 mV/s. Various scan rates are used ranging from 25 mV/s to 150 mV/s with 3 cycles until stationary behavior was reached.[6] Immediately following, 3-cycle cyclic voltammetry (CV) measurements are conducted sweeping from 0 to -4V with a scan rate of 100 mV/s. These voltage limits were used since -3.5V has been identified as the optimal full cell voltage to obtain maximum $CO_2$ products in our system. Optimization was completed iteratively by performing a chronoamperometry (CA) with set voltages ranging from -1.5V to -4V increasing in steps of -0.5V while simultaneously performing gaseous product analysis. Corresponding faradaic efficiency (FE) calculations were performed for each product. After performing the CV measurement, an electrochemical impedance spectroscopy (EIS) measurement was conducted at -1.2V. This measurement assisted in evaluating the relationship between cell resistance and cell performance in addition to membrane selection. Long term stationary electrolysis was then conducted with either a CP or CA at the target potential or current.

Faradaic efficiency calculations were used to evaluate carbon product generation and were computed using gas chromatography sampling results and the recorded current at time of injection. For more information on how FE was calculated, please see supporting documentation. To consider the product transport delay from the cell's cathode output to the gas chromatograph coil, current values were averaged 5 seconds before and after time of injection. These calculations were completed using MATLAB 2016a. An automated plotting and data analysis program was created to ease calculations for lifetime measurements (SI6).

### Results and Discussion:

*Demonstration of stable operation:* Cell design modifications were assisted by evaluating the FE for $CO_2$ and $H_2$ under established $CO_2$ reduction conditions with a Ag electrocatalyst. Since Ag exhibits high selectivity towards CO production, its use is beneficial in cell benchmarking. Lifetime measurements were completed by conducting either a chronoamperometry (CA) or chronopotentiometry (CP) until cell fatality.

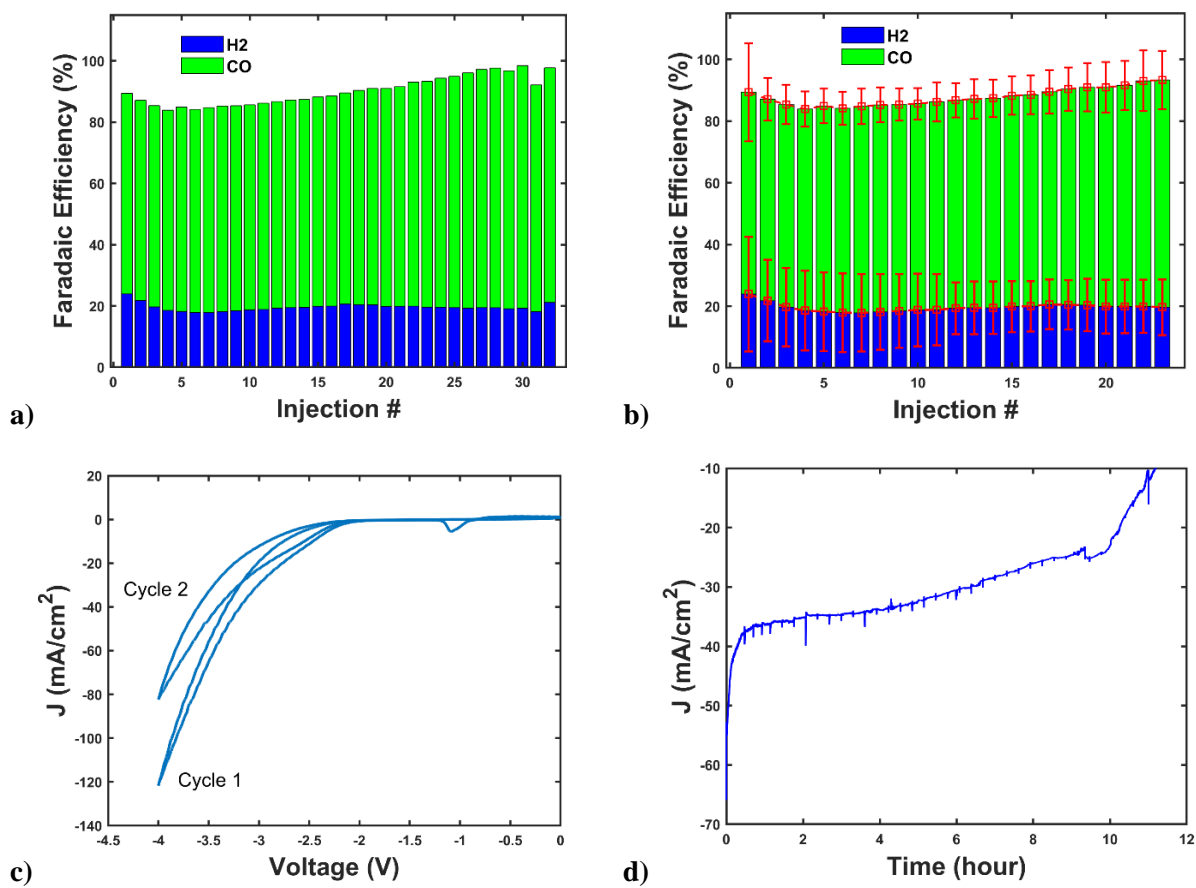

**Figure 3.** CO$_2$ reduction to CO using electrochemical cell configuration shown in Figure 1a and 1b. a) FE (%) for CO and H$_2$ with GDE-Ag electrocatalyst during 12 hours of operation under full cell voltage of -3.5V. b) Variability of FE (%) calculations for three identical experiments during 9 hours of operation. c) Corresponding 3-cycle cyclic voltammograms sweeping from 0 to -4V with scan rate of 100mV/s. d) Resulting current density and CA data under -3.5V full cell voltage.

An average FE of 71.08% for CO is obtained and a maximum FE of 86.03% during 12 hours of operation as shown in Figure 3a with corresponding 30-minute interval GC injections. Cumulative FE for CO and H$_2$ averaged 90.8% and increased to 98.5% at injection 30 out of 33. The loss of cumulative FE may be attributed due to various factors. These may include the current averaging procedure for continuous FE calculations, variance in CO$_2$ feedstock flowrates from liquid contamination in the mass flow controllers, and possible liquid product generation that may crossover through the membrane into the anode and batch anolyte.

The corresponding current density obtained during 12 hours of operation at -3.5V full cell voltage is shown in Figure 3d. The significantly low FE average for CO is due to relatively high current densities obtained during the beginning of electrochemical testing. Highest selectivity for CO is achieved at a range of -30mA/cm$^2$ to -25mA/cm$^2$ and average current density obtained during product analysis was -31mA/cm$^2$. Maximum CO production at current densities of -30mA/cm$^2$ is consistent with other liquid-fed electrolyzers with observed FE degradation occurring at gradually higher current densities.[15,16] The degradation of CO production is proportional to the HER, which is promoted when exposed to high current densities (>30mAh/cm$^2$).

It is observed that with time, selectivity for CO increases while selectivity for H$_2$ production stays relatively constant as shown in Figure 3a and Figure 3b. This may be caused by several factors. First, it may be due to the temporal transport of OH$^-$ ions into the cathode under constant voltage. As OH$^-$ ions diffuse through the membrane, more active sites become available for CO$_2$ on the electrode surface.[18] However, this was only observed for experimental runs where initial current densities obtained during operation were -60mA/cm$^2$ and above. In addition, this may be coupled with the increase of current density under constant voltage, trending toward optimal CO$_2$ reduction current densities.

*Comparison of anode electrode architecture and proof of concept for photocathode incorporation:* Initial design of the electrochemical cell was established with GDE architecture exclusively in the cathode. Since the aim was to combat solubility and performance issues that arise from aqueous systems, a GDE used in the cathode without a catholyte was tested. The anode electrode architecture then consists of a conductive tantalum plate that possesses microwires, sputtered with Ir catalyst on both sides (Figure 1c and 1d). This configuration serves as a flow field to direct reactants from the anolyte solution through the microwires which act as catalyst supports. The microwire configuration also enables incorporation of photocathodes by allowing light absorption through the anode and onto the cathode. This configuration serves as a proof of concept for vapor-fed CO$_2$ (photo)electrochemical reduction even though photocathodes were not used for this experiment.

Long term carbon product analysis was conducted with identical experimental parameters used in demonstrating stable operation above. Therefore, the cathode catalyst was Ag-GDE with the only difference being the anode electrode material. Average current density obtained under -3.5V full cell voltage was -5.7 mA/cm$^2$ during a 9-hour duration (Figure 4a). Maximum FE for CO was 79.33% and average FE obtained for CO was 65.6% (Figure S4). Therefore, with the addition of GDE architecture in the anode, average current densities increase by ~24mA/cm$^2$. Maximum FE for CO increase by 6.7% and average FE obtained for CO increase by 5.2%. Despite operation with the microwire flow field electrode being relatively stable, this was at the cost of current density and faradaic efficiencies for carbon products. In addition, it was observed that with time, the faradaic efficiencies of CO declined after 30 minutes of operation and never fully recovered with subsequent GC injections. In contrast, with GDE architecture in the anode, we show that FE for CO increases with time until cell fatality (Figure 3a, 3d).

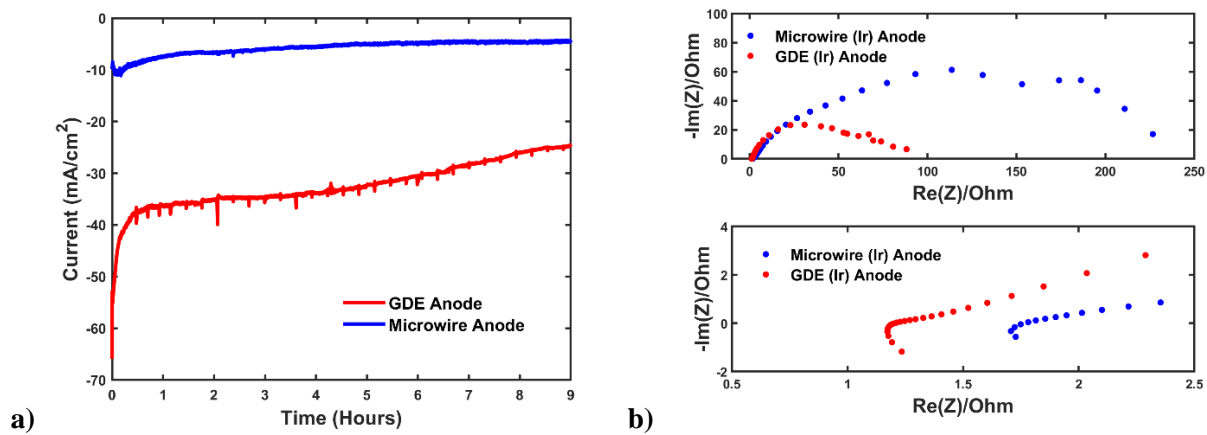

**Figure 4**. EIS measurement under constant DC held at a potential of -1.2V with the reference connected to the counter electrode. a) Comparison of obtained current densities from CA experiment at -3.5V full cell voltage with microwire flow field and carbon based GDE in the anode. b) Top plot: Comparison of PEIS data from both anode configurations. Bottom plot: Zoomed in graphical image showing electrode resistance ($Z_{re}$ value when $-Z_{im}$ is 0).

EIS measurements under constant DC potential of -1.2V are recorded (Figure 4b) for both cell configurations (Figure 1). The cathode electrode-catalyst architecture is identical in both cell configurations with the only difference being the anode electrode component. Comprehensive deciphering of EIS results is beyond the scope of this work. However, obtained EIS measurements elucidate some of the major discrepancies between anode architecture and its effect on overall cell performance (current density and faradaic efficiency). Adapting interpretations of Nyquist plots from work conducted by Mei et al., we observe larger electrode resistance for the tantalum microwire anode than for the GDE anode (Figure 4b). In addition, the largest discrepancy is observed when comparing bulk electrolyte resistance and internal resistance which quantifies the IR drop between electrodes.[21] The bulk electrolyte resistance, which is shown as the diameter of the semicircle in Figure 4b, is 2.5 times larger for the tantalum microwire anode configuration. This could be caused by (i) size differences between the anode and cathode electrode (ii) decrease of reaction sites and active surface area on the microwire (iii) increase of anolyte by volume in the anode compartment.

We can conclude that the use of GDE architecture in the anode decreases cell resistivity and results in increased current densities under operating conditions. We can also conclude that, at the cost of stability, higher selectivity and faradaic efficiencies for reduced carbon products are achieved with GDE architecture. Nevertheless, the use microwire flow fields in the anode compartment demonstrates cell design adaptability and proof of concept for light driven $CO_2$ reduction.

*Demonstration of long-term CO2 reduction:* To demonstrate long term stationary electrolysis of $CO_2$ to carbon products, a CP experiment with a constant current density of -5mA/cm$^2$ was conducted. Electrode architecture is identical to the demonstration of stable operation where FE ~80% for CO was achieved. GC injections were recorded in 15-minute intervals over a 72-hour duration, delivering high resolution carbon product generation data. An industrial grade anion exchange membrane, Selemion was used to validate cell design adaptability, ensuring that proper functionality can be delivered under various operating conditions. Thus, the variance of membranes and electrochemical testing parameters allowed us to solidify the adaptability of our device.

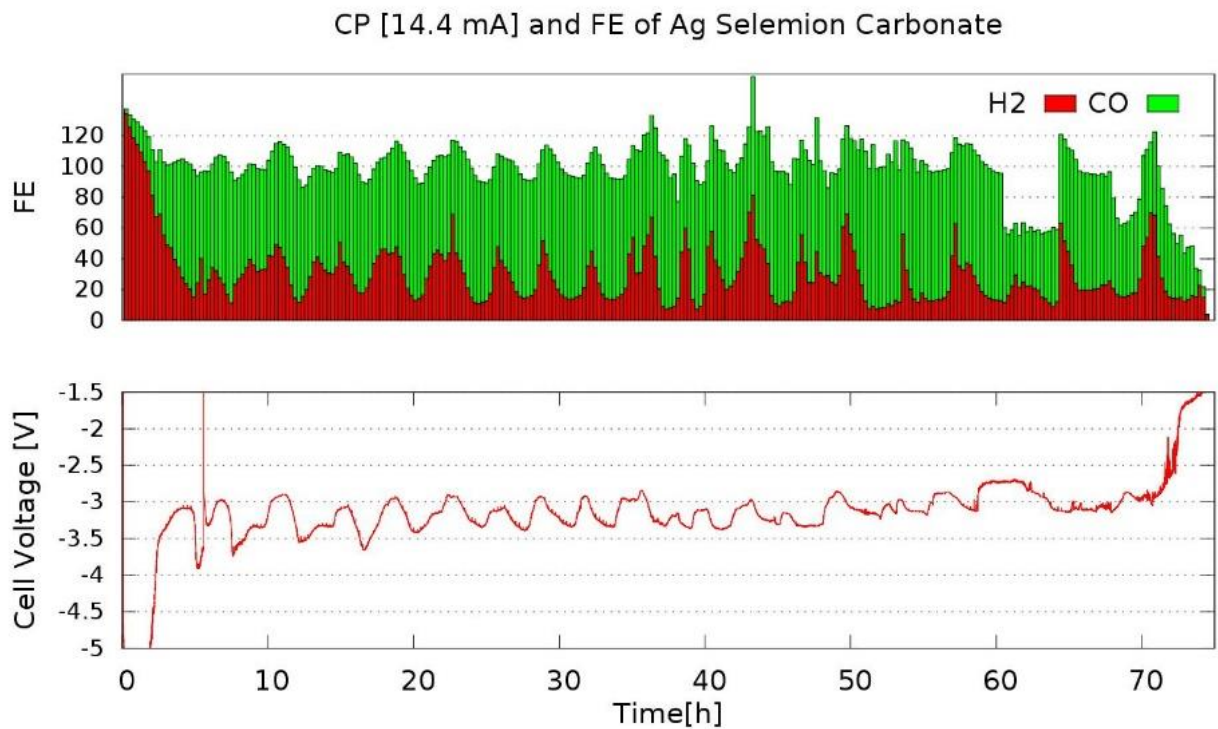

**Figure 5.** FE calculations for CO and $H_2$ during CA experiment at -14.4mA. Plotted below is the corresponding full cell voltage and testing duration in hours.

It is evident that under these operating conditions, cell functionality was temporarily delayed. FE for CO reached 80% after 5 hours of operation. Initial electrochemical products consist mainly of $H_2$, most likely due to the highly negative full cell voltage and resulting large current densities that promote the HER. As the full cell voltage stabilized near -3 to -3.5V, we obtain results consistent with those shown in Figure 3. Represented as FE in Figure 5, CO production gradually increases until a sinusoidal decline consistent with the sinusoidal variation of full cell voltage. The delayed temporal increase of CO production may be attributed to the relatively low applied current density. After 40+ hours of operation, errors are detected in FE calculations with many exceeding 100%. These over-estimations could be due to pressure changes within the cell, affecting feedstock flow rates of $CO_2$.

Despite full cell voltage stability issues with an applied constant current density, our cell design facilitates $CO_2$ reduction under extended operation conditions. A fully vapor fed system, without aqueous support in the cathode, has yet to demonstrate such promising selectivity and efficiency over a 72-hour period.

***Demonstration of selectivity with Cu catalyst:*** To study cell system selectivity for diverse carbon reduction products, a Cu catalyst was used, targeting the high value C-2 product; ethylene ($C_2H_4$).

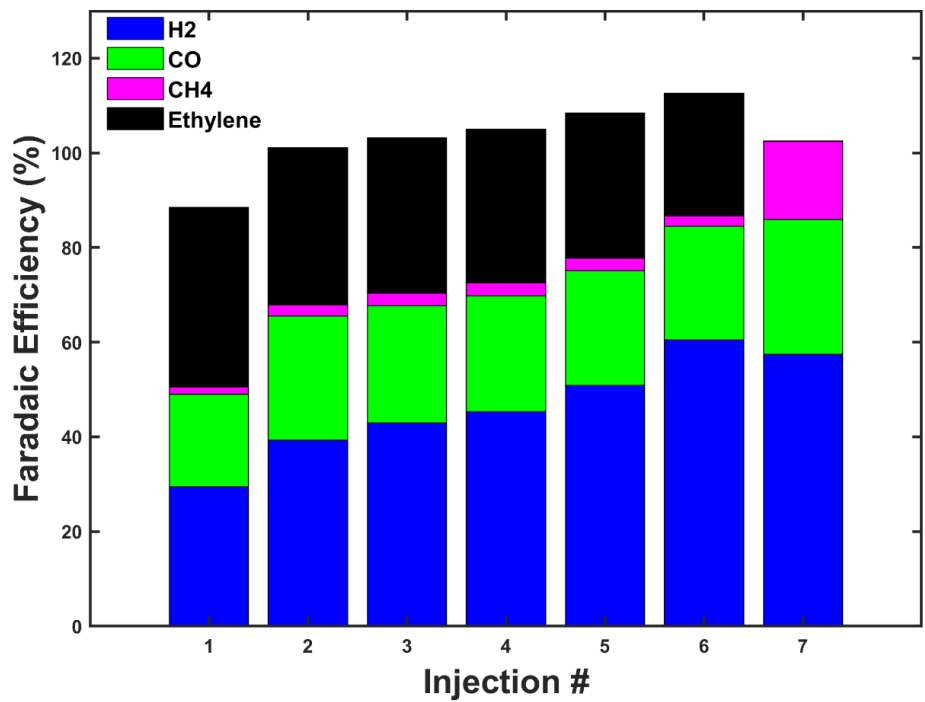

**Figure 6.** Resulting FE for reduced carbon products and hydrogen under constant current density of -17.9mA/cm$^2$.

Under constant applied current of -50mA with resulting current density of -17.9mA/cm$^2$, 40% FE for ethylene was achieved. With subsequent product analysis, a decline in ethylene production is observed. Methane and CO production remain relatively constant until an increase in methane production during GC injection 7, resulting in the absence of ethylene. It is evident that under these conditions, the variable carbon products are $CH_4$, $C_2H_4$, and CO with $H_2$ as a parallel competing reaction. $CO_2$ reduction pathways for $CH_4$ and $C_2H_4$ depend on pH and Cu surface structure with observed elimination dependency between methane and ethylene.[7] The dependence of $CH_4$/$C_2H_4$ production on $HCO_3^-$ concentrations at the electrode surface has been discussed by Hori et al, concluding that $C_2H_4$ production is favored in alkaline conditions while $CH_4$ production is favored in acidic conditions.[23] We experimentally observe simultaneous methane and ethylene production (Figure 6), albeit at low amounts. Thus, the absence of ethylene at injection 7 may be attributed to local changes in pH at the Cu catalyst/membrane interface. Batch anolyte (0.05M $KHCO_3$) was used under constant current. Therefore, initially high $OH^-$ concentrations at the membrane/catalyst interface favored ethylene production, since $OH^-$ species and their electronification has been identified as the main rate determining step.[24] With favored $CH_4$ production at the absence of $C_2H_4$ between GC injection 6 and 7, we can conclude the local pH transitioned towards more acidic conditions. Therefore, increased $C_2H_4$ production and other C-2 products can be achieved in our system if batch anolyte can be held in alkaline conditions.

*Cell fatality and other considerations:* After each experimental run, the cell is disassembled and analyzed. It was evident after long term operation that cell failure was due to the oxidation of carbon based GDEs. This caused the GDEs to disintegrate within the membrane assembly, compromising the catalyst support the carbon paper provided shown in Figure S3. Concerns arose if carbon products were formed from the GDE substrate itself. However, control experiments were conducted with nitrogen as the feedstock and no product generation was observed, showing that our GDE catalyst was not a carbon source (Figure S5).

Direct vapor-fed $CO_2$ architecture does introduce several hurdles in two-electrode electrochemical systems. The absence of a reference electrode hinders potential control of the cell which is vital in systems where the catalyst undergoes aging.[6] Depending on the catalyst and carbon product distributions, the absence of a catholyte also eliminates the ability to perform liquid-product analysis most commonly performed using high-pressure liquid chromatography (HPLC). We conducted HPLC experiments of the cell anolyte after electrochemical experiments and found trace amounts of carbon products. This may have been due to product crossover into the anode after $CO_2$ reduction occurs. Despite these challenges, a fully vapor-fed system with no catholyte simplifies product distribution detection if mainly gaseous products are desired.

## Conclusion:

The cell design and now multi produced prototype has become a testbed for GDE structured electro-catalysts, membranes, and analytes. The design allows for eased catalyst and membrane integration, centered around a zero-gap catalyst/membrane interface and GDE/MEA configuration. We demonstrate the adaptability of our cell design, with the potential for integrating photo absorbers or solar cells for photo enabled $CO_2R$ in the future. Our work confirms the feasibility of vapor-phase $CO_2$ electrochemical reduction under low flow rate conditions, demonstrating stable and extended operation with Ag catalysts and high selectivity for ethylene with a Cu catalyst. This work is a tangible representation of cell design and development that experimentally validates GDE architecture with vapor-fed $CO_2$ conditions. In addition, this experimental work demonstrates cell design feasibility for real world operating conditions and industrial scale up.

**Acknowledgements:**

The authors acknowledge research support from the Joint Center for Artificial Photosynthesis (JCAP), a DOE Energy Innovation Hub, supported through the Office of Science of the U.S. Department of Energy under Award Number DE-SC0004993. Cell design, fabrication, and electrochemical materials development was completed at JCAP. Electrode fabrication was done with tools available at JCAP and the Advanced Light Source (ALS) microfabrication facilities.

**Author Contributions:**

D.L. conceived the idea and designed the cell. A.C. assisted in fabrication and prototyping for proof of concept design. A.C. and D.L. performed electrochemical analysis and gas chromatography. F.M.T. helped design of the experiments, data analysis, and supervised the project.

**References:**

[1] Kim D, Sakimoto K, Hong D, Yang P (2015) *Artificial Photosynthesis for Sustainable Fuel and Chemical Production*. Angew. Chem. Int. Ed. 2015, 54, 2–10.

[2] Andrew B, Bocarsly et al. *Light-driven Heterogenous Reduction of Carbon Dioxide: Photocatalysts and Photoelectrodes.* Chem. Rev. 2015, 115, 12888-12935.

[3] Ridjan I, Mathiesen BV, Connolly D, Duic N (2013) *The feasibility of synthetic fuels in renewable energy systems*. Energy 57:76–84

[4] D.S.A. Simakov, *Renewable Synthetic Fuels and Chemicals from Carbon Dioxide*, SpringerBriefs in Energy, DOI 10.1007/978-3-319-61112-9_2

[5] M. Dry, The Fisher-Tropsch Process: 1955-2000, Catalysis Today 21 (2002) 227-241.

[6] B. Endrődi, G. Bencsik, F. Darvas, R. Jones, K. Rajeshwar, C. Janáky, *Continuous-flow electroreduction of carbon dioxide, Progress in Energy and Combustion Science*, Volume 62, 2017, Pages 133-154.

[7] R. Kortlever et al. *Catalysts and Reaction Pathways for the Electrochemical Reduction of Carbon Dioxide*, J. Phys Chem. Lett. 2015, 6, 4073-4082.

[8] L. Weng, A. T. Bell and A. Z. Weber, Phys. Chem. Chem. Phys., 2018, DOI: 10.1039/C8CP01319E.

[9] Me et al. Carbon nanotubes containing Ag catalyst layers for efficient and selective reduction of carbon dioxide. J. Mater. Chem. A, 2016, 4,8573.

[10] Han et al. High-Rate Electrochemical Reduction of Carbon Monoxide to Ethylene Using Cu-Nanoparticle-Based Gas Diffusion Electrodes. ACS Energy Lett. 2018, 3, 855−860. DOI: 10.1021/acsenergylett.8b00164

[11] Irtem et al. Low-energy formate production from CO2 electroreduction using electrodeposited tin on GDE. J. Mater. Chem. A, 2016, 4,13582

[12] B. Kim et al. Effects of composition of the micro porous layer and the substrate on performance in the electrochemical reduction of CO2 to CO. Journal of Power Sources 312 (2016) 192-198.

[13] El-kharouf et al. Ex-situ characterisation of gas diffusion layers for proton exchange membrane fuel cells. Journal of Power Sources 218 (2012) 393-404.

[14] Merino-Garcia et al. Productivity and Selectivity of Gas-Phase CO2 Electroreduction to Methane at Copper Nanoparticle-Based Electrodes. Energy Technol. 2017, 5, 922-928.

[15] Salvatore et al. Electrolysis of Gaseous CO2 to CO in a Flow Cell with Bipolar Membrane. ACS Energy Lett. 2018, 3, 149-154.

[16] Salvarore et al. Electrolysis of CO2 to syngas in Bipolar Membrane-based Electrochemical Cells. ACS Energy Lett. 2016, 1, 1149−1153

[17] Dinh et al., Science 360, 783–787 (2018) CO2 electroreduction to ethylene via hydroxide-mediated copper catalysis at an abrupt surface.

[18] Kenis et al. The effect of electrolyte composition on the electroreduction of CO2 to CO on Ag based gas diffusion electrodes. Phys.Chem.Chem.Phys., 2016, 18, 7075

[19] Kistler et al. Integrated Membrane-Electrode-Assembly Photoelectrochemical Cell under Various Feed Conditions for Solar Water Splitting. Journal of The Electrochemical Society, 166 (5) H3020-H3028 (2019)

[20] Higgens et al. Gas-Diffusion Electrodes for Carbon Dioxide Reduction: A New Paradigm. ACS Energy Lett. 2019, 4, 317-324.

[21] Mei et al. Physical Interpretations of Nyquist Plots for EDLC Electrodes and Devices. *J. Phys. Chem. C 2018, 122, 194-206.*

[22] Kenis et al. Effect of Cations on the Electrochemical Conversion of CO2 to CO. Journal of The Electrochemical Society, 160 (1) F69-F74 (2013).

[23] Hori, Y.; Murata, A.; Takahashi, R. J. Chem. Soc., Faraday Trans. 1 1989, 85, 2309.

[24] Hori, Y.; Takahashi, R.; Yoshinami, Y.; Murata, A. Electrochemical Reduction of CO at a Copper Electrode. J. Phys. Chem. B 1997, 101, 7075−7081.